\documentstyle[12pt,epsfig]{article}
\newcommand{\be}{\begin{equation}}\newcommand{\ee}{\end{equation}}
\newcommand{\bea}{\begin{eqnarray*}}\newcommand{\eea}{\end{eqnarray*}}
\newcommand{\nn}{\nonumber}
\newcommand{\Real}{\Re e}
\newcommand{\Imm}{\Im m}
\newcommand{\wt}{\widetilde}
\newcommand{\gsim}{\stackrel{>}{_\sim}}

\begin{document}

\thispagestyle{empty}
\renewcommand{\thefootnote}{\fnsymbol{footnote}}
\begin{center}
{\hfill LNF-96/059 (IR)}\vspace{0.2cm} \\
{\hfill hep-ph/9611276}\vspace{1cm} \\
\vglue 1.0 true cm
{\bf {\Large $\eta\to\pi^0\pi^0\gamma\gamma$ to 1-Loop in ChPT}}
\footnote{Work supported in part by HCM, EEC-Contract No. 
CHRX--CT920026 (EURODA$\Phi$NE)}
\vspace{1.0cm} \\
\renewcommand{\thefootnote}{\alph{footnote}}
\setcounter{footnote}{0}
{\bf S. Bellucci}\footnote{E-mail: bellucci@lnf.infn.it}\\
INFN-Laboratori Nazionali di Frascati, P.O. Box 13, 00044 Frascati,
            Italy
\vspace{0.5cm} \\
{\bf Outline} 
\end{center}
1. Introduction: a) motivations, b) status, c) purpose, d) results
\vspace{0.5cm} \\
2. Kinematics and couplings
\vspace{0.5cm} \\
3. Decay amplitude (analytic expression): a) tree level, b) one loop
\vspace{0.5cm} \\
4. Decay width (numerical): a) diphoton energy spectrum\\
b) partial decay rate
[energy cut]
\vspace{0.5cm} \\
5. Discussion: detectability of chiral loop effects\\
(background suppression)
\vfill
\begin{center}
Presented
at the `Workshop on
Hadron Production Cross Sections at DA$\Phi$NE', Karlsruhe, November 1 and 2,
1996
\end{center}
\setcounter{page}0
\renewcommand{\thefootnote}{\arabic{footnote}}
\setcounter{footnote}0
\newpage

\section{Introduction}
                      
a) Motivation:
\vspace{0.5cm} \\
Phenomenological interest for the rare 
decay $\eta\to\pi^0\pi^0\gamma\gamma$ $\Longleftarrow$ Large number of
observed
$\eta$'s anticipated at various $\eta$--factories, e.g. CELSIUS [2
$\times 10^9$], ITEP [$\sim 10^9$],
DA$\Phi$NE [3$\times 10^8~(\phi\to\eta\gamma)$] and other facilities, such as
GRAAL [10$^8$], MAMI, ELSA, CEBAF, ([n]=\#$\eta$'s per year).
\vspace{0.5cm}\\
Theoretical interest: testing chiral perturbation theory (ChPT) (effect of
chiral loops). Of a similar interest [1-9]:
[$\gamma\gamma\to\pi^0\pi^0$,
$\eta\to\pi^0\gamma\gamma$]$=$0 to lowest order (LO)
$\Longrightarrow$ chiral loops are important.
\vspace{1.5cm}\\
b) Status of $\eta\to\pi^0\pi^0\gamma\gamma$:
\bea
A(\eta\to\pi^0\pi^0\gamma\gamma) = A_R + A_{NR}~,\\
\nonumber
\eea
physically distinct.

\noindent
$A_R$ has a pole at $s_{\gamma\gamma}=m_{\pi^0}^2$
($\sqrt{s_{\gamma\gamma}}=$ diphoton invariant mass).
\vspace{0.5cm}\\
$A_R\propto A_{\eta\to 3\pi^0}^{on-shell}A_{\pi^0\to\gamma\gamma}^{on-shell}$
$\Longrightarrow$ Get $A_R$ (up to a phase) from data

\noindent
\hfill (no ChPT calculation needed)
\bea
A_R =  - \frac{A( \eta\to 3 \pi^0 ) A(\pi^0 \to \gamma\gamma )  }{ 
             s_{\gamma\gamma}-m_{\pi^0}^2 }~.    \\
\nonumber
\eea  
$A_R$ dominates over the full kinematical range to LO [10].
\vfill
\newpage
\noindent
$A_{NR}$ must be calculated in ChPT (not from data).
$A_{NR}$ computed at tree level $O(p^4)$ [10]: only $\eta$-exchange diagram
$\eta\to \pi^0\pi^0\eta^* \to \pi^0\pi^0\gamma\gamma$.

\noindent
In [10] also
$\eta^{\prime}$-exchange, formally of higher order (HO).

\noindent
For both diagrams
$A_{NR}$ is negligible with respect to $A_R$, because the LO
$\eta\eta\pi^0\pi^0$ and $\eta\eta'\pi^0\pi^0$
vertices vanish in the limit $m_u=m_d=0$. 
\vspace{0.5cm}\\
Note: analogous suppression factor in the
$\pi^0$--exchange contribution

\noindent
$\propto (m_u-m_d)$,
but thanks to the enhancement due to the pole term
$A_R$ dominates over $A_{NR}$.
\vspace{0.5cm}\\
To one loop?

\noindent
Presumably not: in the (related)
$\gamma\gamma\to\pi^0\pi^0\pi^0$ amplitude
$A_{1-loop}\approx 10A_{tree}$ [11], because
$A_{tree}\propto m_{\pi}^2$ and $A_{1-loop}$ is not suppressed.
\vspace{1.5cm}\\
c) Purpose:

\noindent
to calculate $A_{NR}$ to one loop, neglecting $m_u-m_d$ and
the (suppressed) $\eta$-exchange. Only 1PI diagrams contribute and $A_{NR}$
is finite.

\noindent
$O(p^6)$ counterterms (CT) [12]
do not contribute (as in $\gamma\gamma\to 3\pi^0$).
\vspace{1.5cm}\\
d) Results:

\noindent
$A_{NR}^{1-loop}$ dominates $A_{NR}^{tree}$ (at $m_u=m_d=0$:
$A_{NR}^{1-loop}\neq 0$, $A_{NR}^{tree}=0$).
\vspace{0.5cm}\\
At large $s_{\gamma\gamma}$,
$A_R$ (background for $A_{NR})$ is suppressed $\Longrightarrow$ detect a
pure $O(p^6)$ effect by mesuring $\Gamma(\eta\to\pi^0\pi^0\gamma\gamma)$.
\vfill
\newpage
\noindent

\section{Kinematics and couplings}

a) Kinematics of
$\eta(q) \to \pi^0(p_1) \pi^0(p_2) \gamma(k_1,\epsilon_1)
\gamma(k_2,\epsilon_2)$:
\vspace{0.5cm} \\
Five independent scalar variables:
\bea
&s_{\pi\pi}=(p_1+p_2)^2~, \qquad &z_{1,2}=k_{1,2}\cdot(p_1+p_2)~,\\
\nonumber
&s_{\gamma\gamma}=(k_1+k_2)^2~, \qquad  &z_3=(k_1+k_2)\cdot(p_1-p_2)~.\\
\nonumber
\eea
Decay amplitude:
\bea
A(\eta\to\pi^0\pi^0\gamma\gamma) = 
e^2 \epsilon_1^\mu \epsilon_2^\nu A_{\mu\nu}~.\\
\nonumber
\eea
Decay width:
\bea
\Gamma(\eta\to\pi^0\pi^0\gamma\gamma) = \frac{\alpha^2_{\rm em} }{2^{11}
\pi^6 m_\eta } \int \frac{d^3p_1}{p_1^0} \frac{d^3p_2}{p_2^0}
\frac{d^3k_1}{k_1^0} \frac{d^3k_2}{k_2^0}
\delta^{(4)}(p_1+p_2 + k_1+k_2 ) A^{\mu\nu}A^*_{\mu\nu}~.\\
\nonumber
\eea
\medskip
$A_{\mu\nu}$ is $O(p^4)$

\noindent
(contributions only from odd--intrinsic parity
sector of ChPT $\Longleftarrow$ process involving
the electromagnetic interaction of an odd
number of pions).
\vspace{1.5cm} \\
b) Interaction terms (couplings):
\vspace{0.5cm} \\
$O(p^4)$ ChPT ${\cal L}$:
\bea
{\cal L}={\cal L}^{(2)}+{\cal L}^{(4)}~,\\
\nonumber
\eea
\bea
{\cal L}^{(2)} = {F^2 \over 4 } \mbox{tr}\left( D_\mu U D^\mu U^\dagger
+ \chi U^\dagger  + \chi^\dagger U \right) \\
\nonumber
\eea
\vfill
\newpage
\noindent
${\cal L}^{(4)}$ splits into the odd--intrinsic
anomalous part (i.e. the Wess--Zumino term [13]) and 
the $O(p^4)$ Gasser--Leutwyler lagrangian [14]
\bea
{\cal L}^{(4)}={\cal L}_{WZ}+\sum_{i=1}^{10}L_i{\cal L}^{(4)}_{i}~.\\
\nonumber
\eea
Usual exponential parametrization:
$U=\mbox{exp}(i\sqrt{2}P_8/F)$

\noindent
$P_8 =SU(3)$ 
octet matrix of pseudoscalar mesons

\noindent
$F|_{LO}\equiv\pi^+$ decay constant $F_{\pi}=92.4$~MeV [14,15].
\vspace{0.5cm} \\
Covariant derivative:

\noindent
$D_\mu U = \partial_\mu U +i e A_\mu [Q,U] $

\noindent
$Q=$diag$(2/3,-1/3,-1/3)$.
\vspace{0.5cm} \\
In the external scalar sources: $\chi =\chi^{\dagger}=2 B{\cal M}$

\noindent
${\cal M}=\mbox{diag}(m_u,m_d,m_s)$ quark mass matrix

\noindent
$B|_{LO}\equiv$ mass ratio $B_0=m_{\pi}^2 / (m_u+m_d)$.
\vspace{0.5cm} \\
Neglect in $A_{NR}$
the $\eta-\eta^{'}$ mixing, i.e. mixing of 
$P_8$ with the singlet--field $\eta_0$

\noindent
$\Longrightarrow$ mass--eigenstate $\eta\equiv\eta_8$ octet--field.

\noindent
Also $m_\pi=m_{\pi^0}$ (we neglect isospin--breaking 
in $A_{NR}$).
\vspace{0.5cm} \\
Couplings for tree--level calculation [10]:
\begin{eqnarray*}
A^{(2)}(\eta_8\to \pi^0\pi^0\pi^0) & = &  3 A^{(2)}(\eta_8\to \pi^0\pi^+\pi^-)
= \frac{B_0(m_u-m_d)}{\sqrt{3}F_{\pi}^2}~, \nonumber 
\\
A^{(2)}(\eta_8\to \eta_8 \pi^0\pi^0) & = &  A^{(2)}(\eta_8 \to \eta_8
\pi^+\pi^-) = \frac{B_0(m_u+m_d)}{3 F_{\pi}^2}~, \nonumber
\\
A^{(4)}(\pi^0\to\gamma\gamma ) & = & \sqrt{3} A^{(4)}(\eta_8\to\gamma\gamma ) 
=\frac{e^2}{4\pi^2F_{\pi}} \epsilon_{\mu\nu\alpha\beta}
\epsilon_1^{\mu}k_1^{\nu}\epsilon_2^{\alpha} k_2^{\beta}~.\nonumber
\end{eqnarray*}
\vfill
\newpage
\noindent
Additional couplings for 1-loop diagrams in fig. 1:

\begin{figure}[t]
\centerline{\epsfig{file=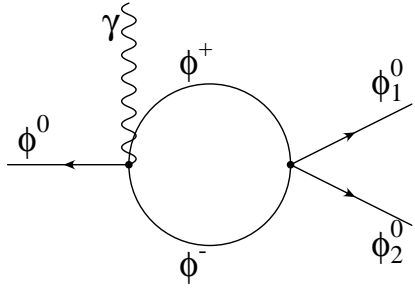,height=4cm}}
\caption{1PI one--loop diagrams for the 
$\phi^0 \to \phi_1^0 \phi_2^0 \gamma\gamma$ transition. 
The second photon line  has to be 
attached to the charged lines running in the loop and to the vertices.} 
 \protect\label{fig1}
\end{figure}
 
\begin{eqnarray*}    
A^{(2)}(\phi^+\phi^-\to\phi_1^0\phi_2^0 ) & = &  a s_{\pi\pi} +
bm_{\pi}^2 + c (p_+^2 -m_{\pi}^2)+d(p_-^2-m_{\pi}^2)~, \nonumber\\ 
A^{(4)}(\phi^0\to\phi^+\phi^-\gamma ) & = & 
f \epsilon_{\mu\nu\alpha\beta}\epsilon^{\mu}k^{\nu}p_+^{\alpha}q^{\beta}~,
\nonumber
\end{eqnarray*}
$q$, $p_{\pm}$, $p_{1,2}$, $k$: (outgoing) momenta of  
the pseudoscalars $\phi^0$, $\phi^{\pm}$, $\phi_{1,2}^0$ and of $A_{\mu}$.

\noindent
$a,b,c$ and $d$ are constants with dim$=m^{-2}$; $[f]=m^{-3}$.
\vspace{0.5cm} \\
$c,d$ are `off--shell couplings' and irrelevant
(they cancel in the amplitude, due to gauge invariance (GI)).

\noindent
In $\pi^+\pi^-\to \pi^0\pi^0$ and $\eta_8\to \pi^+\pi^-\gamma$
we find
\bea
a=-b=\frac{1}{F_{\pi}^2}\qquad\mbox{and} \qquad 
f = -\frac{e}{4\sqrt{3}\pi^2 F_{\pi}^3}~.\\
\nonumber
\eea
(useful to estimate dominant $\pi$-loops).
\vfill
\newpage
\noindent
\setcounter{equation}0
\section{Decay amplitude (analytic)}
a) Tree-level:
\vspace{0.5cm} \\
\bea
A^{(4)}_R=-\frac{e^2}{4\sqrt{3}\pi^2 F_{\pi}^3}
\frac{B_0(m_u-m_d)}{(s_{\gamma\gamma}-m_{\pi^0}^2)}
\epsilon_{\mu\nu\alpha\beta}\epsilon_1^{\mu}k_1^{\nu}\epsilon_2^{\alpha}
k_2^{\beta}~,\label{tree} \\
A^{(4)}_{NR}=-\frac{e^2}{12\sqrt{3}\pi^2 F_{\pi}^3}
\frac{B_0(m_u+m_d)}{(s_{\gamma\gamma}-m_{\eta}^2)}
\epsilon_{\mu\nu\alpha\beta}\epsilon_1^{\mu}k_1^{\nu}\epsilon_2^{\alpha}
k_2^{\beta}~.\label{tree2} 
\eea
$A^{(4)}_R$ enhanced and dominant over $A_{NR}^{(4)}$ in the
entire kinematical space.
\vspace{1.5cm} \\
b) One loop:
\vspace{0.5cm} \\
$O(p^6)$ loop and CT divided in 
three GI subgroups: reducible $\pi^0$--exchange diagrams, reducible 
$\eta_8$--exchange diagrams and 1PI diagrams. 
\begin{enumerate}

\item[i.] $\pi^0$--exchange diagrams (include both loops and CT) 
contribute mainly to $A_{R}$.

\noindent
In principle they contribute also 
to $A_{NR}$. Decompose the $\eta\to \pi^0
\pi^0(\pi^0)^*$ 
amplitude:
\bea
A(\eta\to \pi^0 \pi^0(\pi^0)^*)=A_{on-shell}(\eta\to 3\pi^0)
+  (s_{\gamma\gamma}-m_{\pi}^2)\times A_{off-shell}~.\\
\nonumber
\eea
Non--resonant contribution $\propto {\cal A}_{off-shell}$
vanishes in the limit $m_u=m_d$ $\Longrightarrow$
neglected.

\noindent
Extract $|{\cal A}_{on-shell}|$ from experiments, no need to
evaluate it in ChPT.
\vfill
\newpage

\item[ii.] $\eta_8$--exchange diagrams (both loops and CT)
contribute only to $A_{NR}$ and can be 
neglected.

\noindent
We explicitly  checked that 
they are same order as
tree--level $A^{(4)}_{NR}$ (small).

\noindent
Reason of suppression:
$\pi$-$\pi$ loops (dominant  
contribution) are suppressed by $(m_u+m_d)$
(as the tree level).

\noindent
$K$-$K$ loops and ${\cal L}^{(4)}$
are not suppressed 
by $(m_u+m_d)$. Nonetheless negligible

\noindent
(we
are far below the kaon threshold and the CT combinations 
involved, i.e.~  $(L_1+L_3/6)$, $(L_2+L_3/3)$ and $L_4$,  
are small [16]). 

\item[iii.] The 1PI diagrams: fig.~1 (at least four distinct diagrams).
\noindent

Their sum is finite and is the dominant contribution to $A_{NR}$.   

\end{enumerate}
\vfill
\newpage
\noindent
Calculation of loop diagrams in fig.~1 similar to 
[17]:
the radiative four--meson  amplitudes, with
one pseudoscalar replaced by one photon (difference).

\noindent
Results simply dictated by QED
\bea
A_{NR}^{1PI} &=& 4 e f  (as_{\pi\pi}+bm_\pi^2) \times \nn\\
&&\!\!\!\! \times \Biggr\{ \widetilde{ C_{20}}( s_{\pi\pi},-z_2) 
\epsilon_{\mu\nu\alpha\beta}
\epsilon_1^{\mu}k_1^{\nu} \left[(\epsilon_2 \cdot p_{12}) k_2^{\alpha}
 - z_2 \epsilon_2^\alpha
\right] q^{\beta}\; +\; (\epsilon_1,k_1 \leftrightarrow \epsilon_2,
k_2) \Biggr\}~, \label{1-l}
\eea
$p_{12}=p_1+p_2$
\vspace{0.5cm} \\
function $\wt{C_{20}}(x,y)$ defined
in terms of the 
three--denominator one--loop scalar functions [17].

\noindent
In $\pi$-$\pi$ case and for $x,\ x-2y > 4 m_\pi^2$ the 
explicit expression is:
\bea
(4\pi)^2 \Real \wt{C_{20}}(x,y) &=& {x \over 8 y^2}\left\lbrace 
\left(1-2{y\over x}\right)\left[\beta\log\left({1+\beta
\over 1-\beta}\right) -\beta_0\log\left({1+\beta_0
\over 1-\beta_0}\right)\right] \right. \nn\\ 
&& \left. \qquad + 
{m_\pi^2\over x} \left[\log^2\left({1+\beta_0 \over 1-\beta_0}
\right)-\log^2\left({1+\beta \over 1-\beta }\right) \right]
+ 2{y\over x} \right\rbrace~, \\
& & \nn \\
 (16\pi) \Imm \wt{C_{20}}(x,y) &=& - {x \over 8 y^2}\left\lbrace 
\left(1-2{y\over x}\right)\left[\beta-\beta_0\right] \right. \nn\\ 
&& \left. \qquad + 
{2 m_\pi^2 \over x} \left[\log\left({1+\beta_0 \over 1-\beta_0}
\right)-\log\left({1+\beta \over 1-\beta }\right) \right]
+ 2{y\over x} \right\rbrace~,\\
\nonumber
\eea
\bea
\mbox{where} \qquad
\beta_0=\sqrt{1-{4m_\pi^2\over x }}\qquad\mbox{and}\qquad
\beta=\sqrt{1-{4m_\pi^2\over ( x-2y )}}~.
\eea
Due to GI, amplitude depends
only on `on--shell couplings' $a$, $b$ and $f$. 

\noindent
Result is $O(k_1,k_2)$ (analogy to $O(k)$ direct--emission amplitudes
of [17]).
\vfill
\newpage
\noindent
Vertices in a general form $\Longrightarrow$ not only dominant pion loops,
but also kaon loops are represented in result.

\noindent
We recover, as a particular
case, part of the result of [11] (i.e. the
1PI diagrams).

\noindent
Correspondence of
$\widetilde{ C_{20}}(x,y)$ with their
function $R(x,y)$:
\bea
R(x,y)=32\pi^2 y \wt{ C_{20}}(x,y)~. \label{RC}
\eea
Result depends only on $\widetilde{ C_{20}}$ and thus is finite
$\Longleftarrow$
\begin{enumerate}

\item GI of the amplitude and

\item on--shell $\pi^+\pi^-\to\pi^0\pi^0$ amplitude independent of 
loop variables (it depends only on $s_{\pi\pi}$).

\end{enumerate}
Sum of 1PI diagrams is no more finite
if the two external $\pi^0$'s are replaced by a $\pi^+$-$\pi^-$ pair
$\Longleftarrow$
\begin{enumerate}

\item on--shell
$\pi^+\pi^-\to\pi^+\pi^-$ amplitude depends on loop momenta,
also

\item sum is not GI (to get GI result, add
reducible diagrams with a photon emission from external legs).    

\end{enumerate}
\vfill
\newpage
\noindent

\setcounter{equation}0
\section{Decay width (numerical)}
\begin{figure}[t]  
   \begin{center}
   \setlength{\unitlength}{1truecm}
       \begin{picture}(10.0,10.0)
       \epsfxsize 10.0 true  cm
       \epsfysize 10.0 true cm
       \epsffile{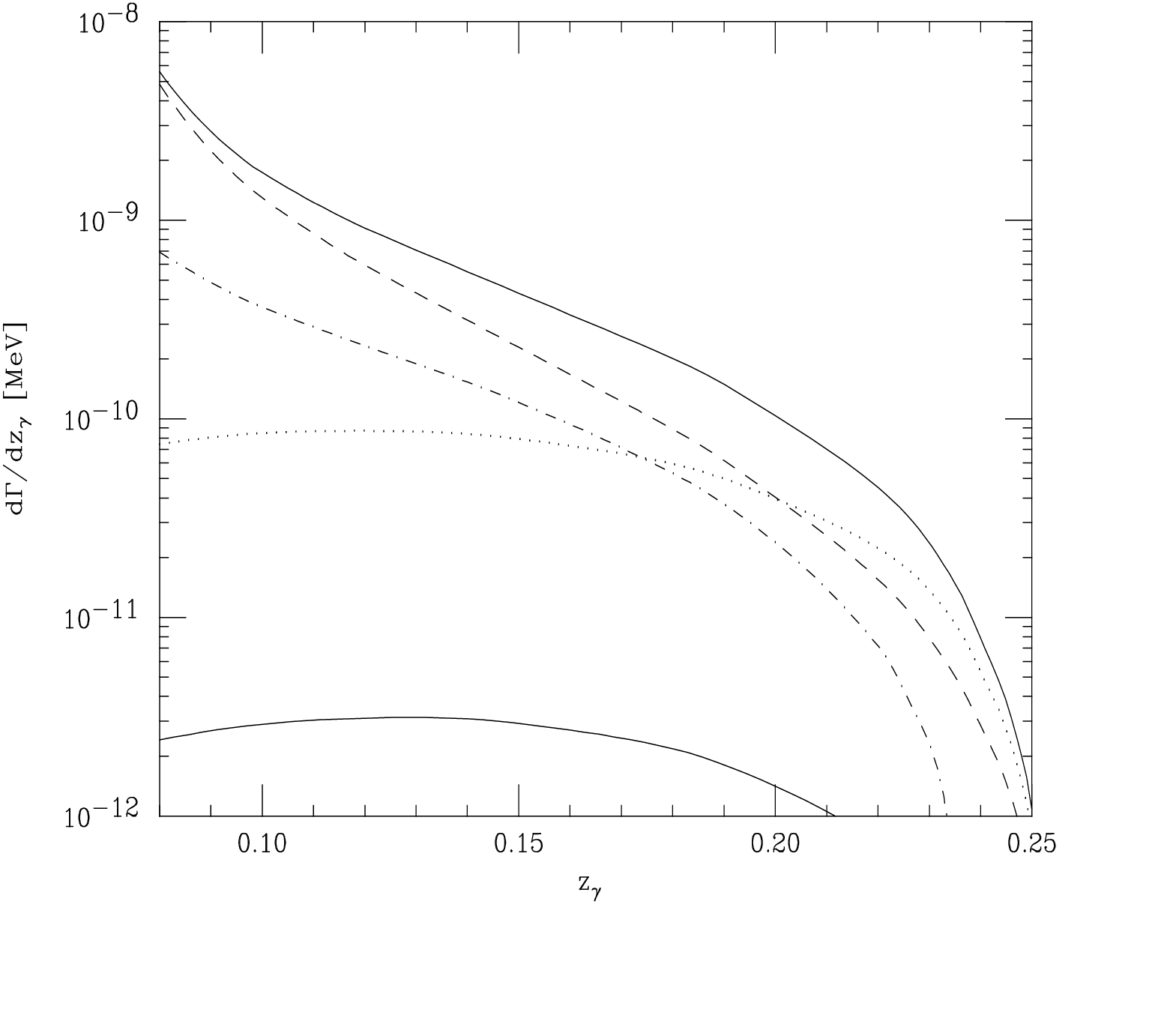}
   \end{picture}
   \end{center}
   \caption{Diphoton spectrum ($z_\gamma=s_{\gamma\gamma}/m_\eta^2$)
for the decay $\eta\to\pi^0\pi^0\gamma\gamma$. The upper full line is
the total contribution. The dashed line is the resonant contribution   
($|A^{phys}_R|^2$), the dotted line is the one--loop non-resonant 
contribution ($|A^{1PI}_{NR}|^2$) and the 
dash--dotted line is their interference ($\rho=2,~\alpha_0=0.18$). The 
lower full line is the tree-level non--resonant contribution 
($|A^{(4)}_{NR}|^2$).} 
   \protect\label{fig2}
\end{figure}
 
\noindent
Results of numerical analysis: figs. 2 and 3 from [18].

\noindent
Plots obtained integrating numerically 
the (modulus-square of) decay amplitude:
\bea
A(\eta\to\pi^0\pi^0\gamma\gamma) = A_R^{phys} +\left[ A^{(4)}_{NR}
+A^{1PI}_{NR} \right]~.  \\
\nonumber
\eea
$A^{(4)}_{NR}$, 
$A^{1PI}_{NR}$ are the ChPT results.
\vspace{0.5cm}\\
$A_R^{phys}$ is a phenomenological expression for
the resonant amplitude:
\bea
A^{phys}_R=  A^{(4)}_R  \rho e^{i\alpha_0}~.  \label{Aphys} \\
\nonumber
\eea
Factor $\rho e^{i\alpha_0}=$ corrections to tree--level
$\eta\to 3\pi^0$ amplitude (known to be large [19]).
\vspace{0.5cm}\\
$\rho$ can be obtained from data:

\noindent
assume a flat Dalitz Plot for $\eta\to 3\pi^0$ decay

\noindent
(no experimental evidence of a D-wave contribution)

\noindent
and use [19]
\bea
B_0(m_u-m_d)= m^2_{K^0} - m^2_{K^+} - m^2_{\pi^0} + m^2_{\pi^+}~,\\
\nonumber
\eea
$\Longrightarrow$ we find $\rho \simeq 2$.
\vfill
\newpage
\noindent
Phase $\alpha_0$ cannot be extracted from $\eta\to 3\pi^0$ data.
\vspace{0.5cm}\\
Evaluate $\alpha_0$, similarly to $K\to 3\pi$ analysis of
[20]:

\noindent
expand the one--loop 
$\eta\to 3\pi^0$ amplitude of [19]
around the center of the Dalitz Plot 
\bea
\Longrightarrow
\alpha_0 = \frac{ 1}{32 \pi F_{\pi}^2} \left( 1 - \frac{4 m_\pi^2 }{s_0} 
\right)^{1/2} (2 s_0 + m_\pi^2) \simeq 0.18~,\\
\nonumber
\eea
where $s_0=(m_\eta^2 +3m_\pi^2)/3$.
\vspace{0.5cm} \\
\begin{figure}[t]  
   \begin{center}
   \setlength{\unitlength}{1truecm}
       \begin{picture}(10.0,10.0)
       \epsfxsize 10.0 true  cm
       \epsfysize 10.0 true cm
       \epsffile{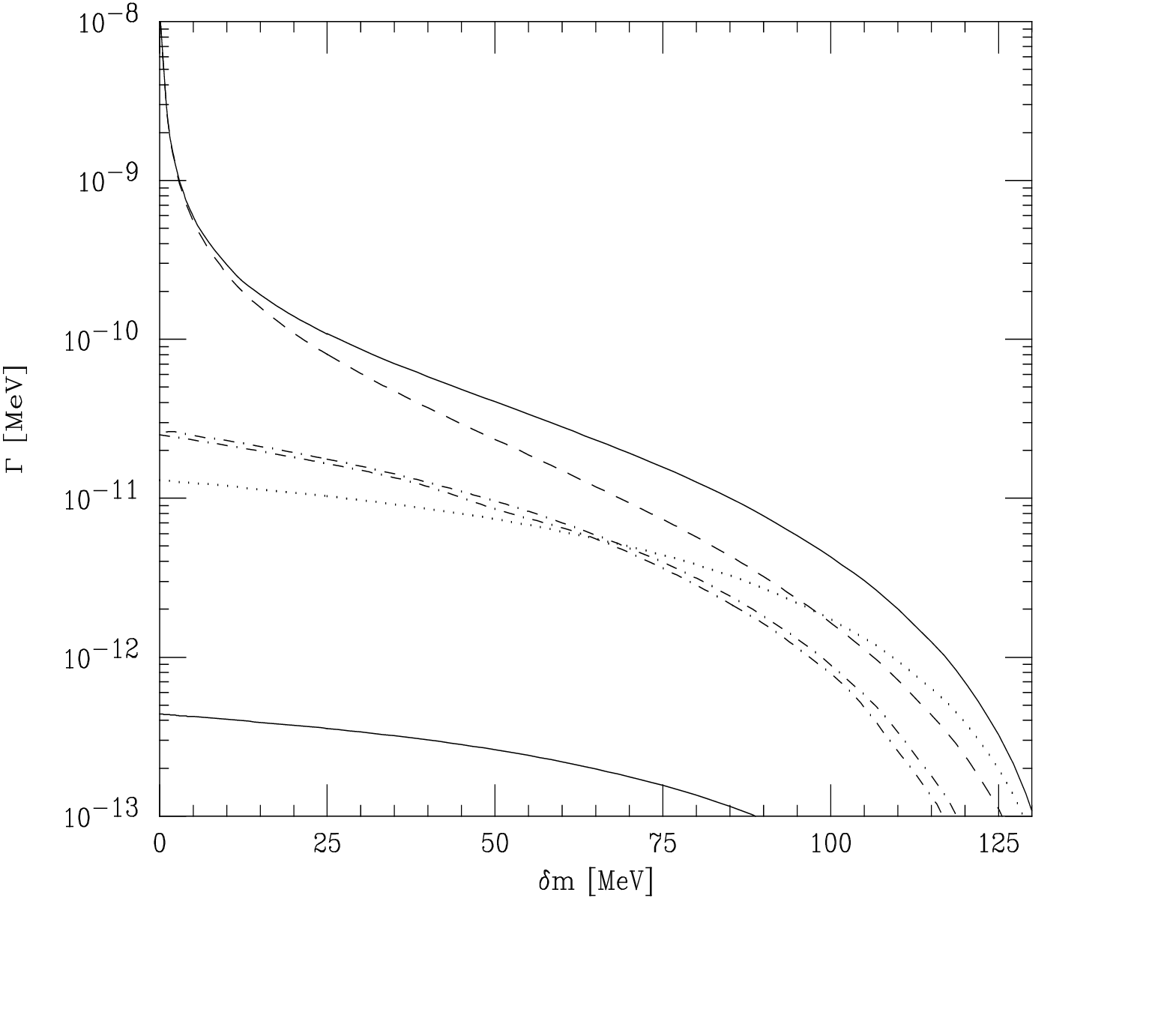}
   \end{picture}
   \end{center}
   \caption{Partial decay rate of 
$\eta\to\pi^0\pi^0\gamma\gamma$ 
as a function of the energy cut  
$|s_{\gamma\gamma}^{1/2}-m_{\pi^0}|<\delta m$. 
Full, dashed and dotted curves as in fig.~2. The two dash--dotted lines,
denoting the interference between $A^{phys}_R$ and $A^{1PI}_{NR}$, 
have been obtained for $\alpha_0=0.16$ (upper line) and 
$\alpha_0=0.20$ (lower line).}
   \protect\label{fig3}
\end{figure}
Figs. 2,3 show that:
\begin{enumerate}

\item $A_{NR}^{1PI}$ dominates over $A_{NR}^{(4)}$ in the
whole phase space.

\item For $s_{\gamma\gamma} \gsim 0.15 m^2_\eta$: $A_{NR}$
becomes non--negligible with respect to $A_R$.

\item For $s_{\gamma\gamma} \gsim 0.20 m^2_\eta$: $A_{NR}$ gives
the dominant contribution.

\end{enumerate}
Used $\rho=2$ in $A_R^{phys}$ and dominant $\pi$-$\pi$ loops only
in $A_{NR}^{1PI}$. 

\noindent
Kaon loops  give a very small contribution (checked). 
\vspace{0.5cm}\\
Fig 3: result is quite independent of $\alpha_0$.

\noindent
Normalization factor $\rho$ is very important. 
\vspace{0.5cm}\\
More precise data on  $\eta\to 3\pi^0$ 
$\Longrightarrow$ improve
the accuracy on $A_R^{phys}$
$\Longrightarrow$ include the (small)
$D$--wave contribution we neglected.
\vspace{0.5cm}\\
Discrepancy with [10] in overall normalization 
factor.
\noindent
Analytic agreement. Problem in the program 
used to produce [10] plots.
\vfill
\newpage
\noindent
At DA$\Phi$NE, assuming luminosity ${\cal L}=5\times 10^{32}$ cm$^{-2}$s$^{-1}$,
1 year $=10^7$ s

\noindent
$\Longrightarrow$ \# of $\phi\to\eta\gamma$ decays per year
$=2.8\times 10^8$.

\begin{table}
\[ \begin{array}{|c|c|c|c|c|} \hline
\delta m[\mbox{MeV}]  & 0   & 25    & 50  & 75   \\ \hline
    \mbox{Br}    & 0.3 &  10^{-7} & 3\times 10^{-8} &  10^{-8} 
  \\ \hline 
 \mbox{N/year}  &  9\times 10^{7} &  30 &  9 &  3 \\ \hline
\Gamma_{NR}/\Gamma_{R} &  -  &  0.4 &  1 &  1.5 \\ \hline   
 \end{array}    \]
\end{table}

\noindent
Used: $\Gamma_{tot}(\eta)=1.18\times 10^{-3}$ MeV, Br$(\eta\to 3\pi^0)=32.1\% $

\noindent
[recalling: Br$(\pi^0\to\gamma\gamma)=99\% $]
\vspace{0.5cm} \\
$\Longrightarrow$ total \# of $(\eta\to 3\pi^0)$
events (no cut, i.e. $\delta m=0$) $=9\times 10^7$ per year.
\vspace{1.5cm} \\
\begin{eqnarray}
\Gamma_R &=& \int d\Gamma |A_R|^2  \nonumber \\
\Gamma_{NR} &=& \int d\Gamma \Biggr( |A_{NR}|^2
+2Re(A_R^{*}A_{NR})\Biggr) \nonumber
\end{eqnarray}
\vfill
\newpage
\noindent

\setcounter{equation}0
\section{Discussion: detectability of chiral loop effects
(background suppression)}

Dominant 1-loop corrections in ChPT to $\eta\to\pi^0\pi^0\gamma\gamma$,
to go beyond the simple current algebra calculation of [10].
\vspace{0.5cm}\\
Phenomenological interest:
experimental facilities
acting effectively as $\eta$-factories.
\vspace{0.5cm}\\
Results on $\gamma\gamma\to\pi^0\pi^0\pi^0$ [11]
inspiring:
\begin{enumerate}

\item lowest--order
amplitude is suppressed and

\item the corrections due to chiral loops
dominate the cross-section.

\end{enumerate}
\renewcommand{\arraystretch}{1.2}
\parskip=4pt plus 1pt
Similar result for the non--resonant contribution
to $\eta\to\pi^0\pi^0\gamma\gamma$.
\vspace{0.5cm}\\
Despite this enhancement (due to 1-loop corrections),
$A_{NR}$ is shadowed from $A_R$ (i.e. the 
$\pi^0$--exchange) over
a large portion of the diphoton spectrum.
\vspace{0.5cm}\\
However, at large
$s_{\gamma\gamma}$, $A_{NR}^{1PI}$ dominates over $A_R$.
\vspace{0.5cm}\\
Measurement of $\eta\to\pi^0\pi^0\gamma\gamma$ partial width
in this region $\Longrightarrow$
new test of ChPT at $O(p^6)$.
\vfill
\newpage
\noindent
Future developments.
\vspace{0.5cm}\\
$\eta\to\pi^+\pi^-\gamma\gamma$ (statistically favored):
dominated
by the bremsstrahlung of  $\eta\to\pi^+\pi^-\gamma$ [10]

\noindent
(not suppressed already at the tree level)
$\Longrightarrow$
1-loop corrections not related to the  $\eta\to\pi^+\pi^-\gamma$ 
amplitude will be hardly detectable.
\vspace{0.5cm}\\
$\gamma\gamma\to\pi^+\pi^-\eta$
and $\gamma\gamma\to\pi^0\pi^0\eta$ more interesting
for studying chiral--loop effects.
\newpage
\section*{Acknowledgments}
I wish to thank W. Kluge for the invitation
to this workshop.
I also acknowledge my collaborator G. Isidori for valuable help
in the preparation of this talk.

\end{document}